\documentclass[conference]{IEEEtran}
\IEEEoverridecommandlockouts
\usepackage{cite}
\usepackage{amsmath,amssymb,amsfonts}
\usepackage{algorithmic}
\usepackage{graphicx}
\usepackage{textcomp}
\usepackage{xcolor}

\newcommand{\Nf}{$N_\mathrm{filter}$}

\ifCLASSOPTIONcompsoc
    \usepackage[caption=false, font=normalsize, labelfont=sf, textfont=sf]{subfig}
\else
\usepackage[caption=false, font=footnotesize]{subfig}
\fi

\def\BibTeX{{\rm B\kern-.05em{\sc i\kern-.025em b}\kern-.08em
    T\kern-.1667em\lower.7ex\hbox{E}\kern-.125emX}}
\begin{document}

\title{Low Complexity Channel estimation with Neural Network Solutions}

\author{\IEEEauthorblockN{Dianxin Luan, John Thompson}
\IEEEauthorblockA{\textit{Institute for Digital Communications, School of Engineering, University of Edinburgh, Edinburgh, EH9 3JL, UK.}\\
Email address : Dianxin.Luan@ed.ac.uk, john.thompson@ed.ac.uk}
}

\maketitle

\begin{abstract}
% JST 29/07: some minor tweaks to the abstract
Research on machine learning for channel estimation, especially neural network solutions for wireless communications, is attracting significant current interest. This is because conventional methods cannot meet the present demands of the high speed communication. In the paper, we deploy a general residual convolutional neural network to achieve channel estimation for the orthogonal frequency-division multiplexing (OFDM) signals in a downlink scenario.
Our method also deploys a simple interpolation layer to replace the transposed convolutional layer used in other networks to reduce the computation cost. The proposed method is more easily adapted to different pilot patterns and packet sizes.
Compared with other deep learning methods for channel estimation, our results for 3GPP channel models suggest improved mean squared error performance for our approach. 
\end{abstract}

\begin{IEEEkeywords}
Channel estimation, Residual convolutional neural network, Orthogonal frequency-division multiplexing (OFDM)
\end{IEEEkeywords}

\section{Introduction}

To achieve high data rate transmission to meet the demand of the modern wireless applications, wireless communication systems are explored for high capacity. For fourth and fifth generation (5G) communication systems, orthogonal frequency-division multiplexing (OFDM) is widely deployed as the baseband modulator to achieve  high speed communication. It provides the superior performance through its resistance to frequency-selective channel fading and high bandwidth efficiency by frequency-division multiplexing. However, precise channel state information also needs to be obtained. Conventional channel estimation methods include least squares (LS) and minimum mean squared error (MMSE) \cite{b1}. LS is a simple approach, while MMSE can provide an improved solution, but which requires accurate knowledge of channel statistics. That becomes challenging for conventional methods when high data rates are required and the operation environments are very complex. 

With the success of the neural network applications in computer vision \cite{b2} and natural language processing \cite{b3}, it motives researchers to explore the neural network, especially deep learning (DL) based channel estimation methods \cite{b4} \cite{b5}. Compared with  conventional methods which aim to find analytical solutions, neural network methods are a type of data driven model. Neural network methods, which update the parameters by the back-propagation algorithm \cite{b6}, give us an alternative methodology for system design. This method can be useful given that communication systems are often implemented in very complex environments and closed form solutions may not exist. Compared with LS and MMSE methods, neural network solutions can offer competitive performance and do not exploit prior knowledge of channel statistics necessarily but need to be trained on example channels offline. Although the estimated MMSE method \cite{b} is also an powerful option to achieve good performance in practice, it will perform poorly if the mathematical model for the channel is uncertain. In particular, if the assumed power delay profile does not match the propagation channel model, the performance of estimated MMSE methods can be significantly degraded. That motivates the implementation of neural network methods, especially DL, as an alternative solution to handle the channel estimation task. It is important to focus on the generalization capability of neural network to generalize to perform well with the practical channel models. 

% JST updated last sentence of this paragraph
As a result, channel estimation with deep learning has recently been widely investigated. Two recently published channel estimation networks are ChannelNet \cite{b5} and ReEsNet \cite{b7}. ChannelNet presents a deep learning algorithm for channel estimation in communication systems. The time–frequency response of a fast fading communication channel is considered as a two dimensional image. The aim is to find the unknown values of the channel response given the known values at the pilot locations. The chosen DL design cascades a super-resolution network and an image restoration method. Compared with ChannelNet, ReEsNet is a residual convolutional neural network which can outperform the ChannelNet and the computational cost is also lower. It should be noted that the gradient vanishing and the degeneration problems in DL are mitigated by using a residual neural network. However, the hyperparameters of ReEsNet need to be adapted to obtain the best performance for different pilot patterns.

Inspired by that, in this paper, we introduce a low complexity neural network solution for channel estimation which can achieve satisfactory estimation performance for high data rate communication links. The proposed method is flexible for any pilot pattern in the downlink scenario and has acceptable generalization capability for different channel models defined in the 3GPP standardization documents. The remainder of this paper is organised as follows. Section \ref{Conventional methods for channel estimation} introduces conventional methods implemented for channel estimation. Section \ref{Neural Network Methods} discusses the Channelnet and ReEsNet, then proposes our neural network solution, called Interpolation-ResNet. Section \ref{Simulation results} shows the simulation results of the individual performance and generalization. Section \ref{Conclusion} summarizes the key points of the paper. 

\section{Conventional methods for channel estimation}
\label{Conventional methods for channel estimation}

\begin{figure}[htbp]
\centerline{\includegraphics[width=0.35\textwidth]{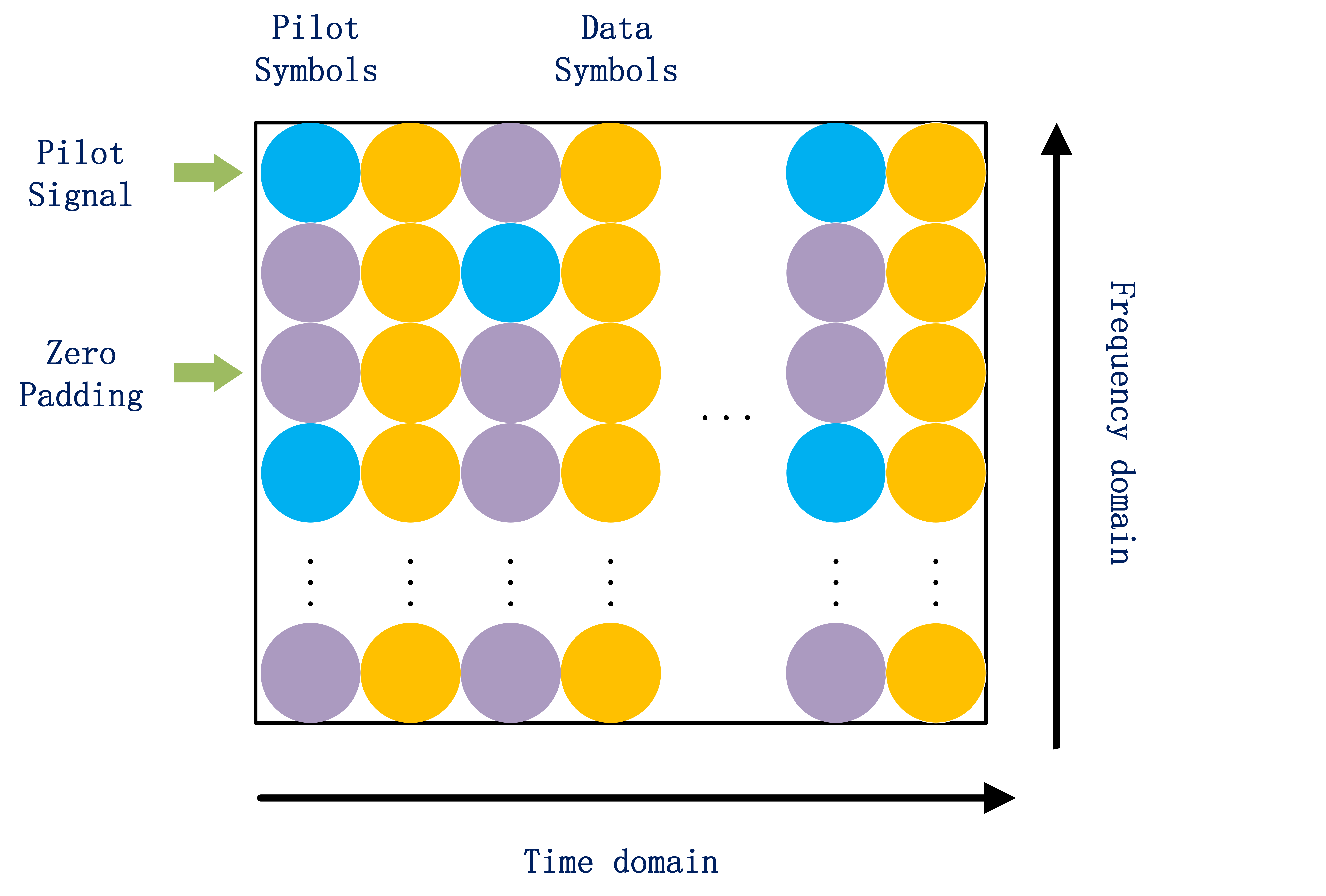}}
\caption{Pilot and data symbols of one frame}
\label{Frame Diagram}
\end{figure}

Considering that OFDM is implemented as the baseband modulator in the single-input and single-output (SISO) downlink scenario, let $x(t)$ denote the transmitted signal in the time domain. Then, the received signal $y(t)$ in the time domain is given by
% JST 29/07 - I have removed spaces between the text and equations to save space!
\begin{equation}
y(t) = g(t)\otimes x(t) + n(t)
\end{equation}
Where $n(t)$ denotes additive white Gaussian noise (AWGN) and $g(t)$ is the impulse response of the Rayleigh fading channel. The channel impulse response is assumed to vary slowly over each frame, which consists of $N_t$ OFDM symbols. If the maximum path delay is less than the duration of Cyclic Prefix (CP), the equivalent frequency domain expression for the $n$th OFDM symbol will be given by
\begin{equation}
Y(n) = H(n) \circ X(n) + N(n)
\end{equation}
Where $H(n), X(n), N(n) \in \mathbb{C}^{N_f}$ denote the discrete Fourier Transforms of $g$, $x$ and $n$ respectively, which contains the channel gain, the current frequency domain symbols being transmitted and the AWGN noise present on the $N_f$ subcarriers. $N_f$ is the length of the fast Fourier Transform (FFT) used in the OFDM receiver and $\circ$ denotes the element-wise multiplication operation. Fig.~\ref{Frame Diagram} shows the frame structure including the pilot OFDM symbols and the data OFDM symbols. For the data OFDM symbols, all the subcarriers transmit modulated signals. For the pilot OFDM symbols, the selected subcarriers are used for pilot and the other subcarriers are set to 0, in a similar manner to the 5G demodulation reference signals \cite{b12}. 
The received sparse pilots are extracted and exploited for frequency domain channel estimation to predict the channel gains for the whole frame. That means that channel estimation methods use the known pilot signal to obtain the predicted channel gain $\hat{H} \in \mathbb{C}^{N_f \times N_t}$ as a precursor to demodulating the distorted data symbols. The conventional channel estimation methods are the least squares (LS) and minimum mean square error (MMSE) approaches which are now introduced. 

\subsection{LS method}

By minimizing the Euclidean distance of $Y$ and $H \circ X$, the estimated channel gain matrix is written as follows.
\begin{equation}
    \hat{H}_{LS} = \frac{Y_{Pilot}}{X_{Pilot}}
\end{equation}
Where $Y_{Pilot}, X_{Pilot}$ denote the vector of received and transmitted pilot signals respectively. LS aims to minimize the error between the vectors $Y$ and $H \circ X$ but the LS operation can lead to amplification of the noise level in the estimation $\hat{H}_{LS}$ \cite{b8}. The predicted channel gains of the whole frame is obtained by interpolating $\hat{H}_{LS}$ linearly in both the time and frequency domain.

\subsection{MMSE method}

By minimizing the Euclidean distance between $H$ and $H_{LS}$, linear MMSE estimation \cite{b9} of the channel at the pilot OFDM symbol is written as follows: 
% JST 29/07 - I have updated the notation to represent the channel parameters directly
\begin{equation}
    \hat{H}_{MMSE} = R_{HH_{p}}\left(R_{H_{p}H_{p}} + I\frac{\sigma_N^2}{\sigma_X^2}\right)^{-1}\hat{H}_{LS}\\
\end{equation}
Where $H$ is the channel gain matrix at the pilot symbol, $H_p$ denotes the real measured channel gain matrix for the pilot subcarriers and ${\sigma_N^2}/{\sigma_X^2}$ is the numerical reciprocal of the signal-to-noise ratio (SNR). The scalars $\sigma_N^2$ and $\sigma_X^2$ denote the average power of the transmitted signal and the AWGN noise respectively. $R_{HH_p}$ and $R_{H_pH_p}$ are the crosscorrelation matrix of $H, H_{p}$ and the autocorrelation matrix of $H_{p}$, which are defined as:
\begin{equation*}
\begin{split}
\label{correlation}
    R_{HH_p} = E\{HH_{p}^{H}\} , \hspace{0.5cm}
    R_{H_pH_p} = E\{H_pH_{p}^{H}\}
\end{split}
\end{equation*}
MMSE estimation enhances the performance of the LS method by accessing the prior knowledge from the channel, which unfortunately is not a causal operation. In practice, these correlation matrices may be estimated from past channel measurements obtained from previous pilot signals. To predict the channel matrix for the whole frame, linear interpolation is deployed to resize the estimated channel matrix $\hat{H}_{MMSE}$. 

\section{Neural Network Methods}
\label{Neural Network Methods}

% JST: updated last sentence
Neural network methods are attracting more and more interest to provide superior performance compared with the conventional methods described in Section II \cite{a}. However, the conventional neural network solution can have problems with gradient vanishing issues, so the residual neural network can be deployed to reduce these effects. We will discuss the state-of-the-art ChannelNet and ReEsNet networks and then introduce our proposed Interpolation-ResNet network, which can perform better than these existing solutions.

\subsection{ChannelNet and ReEsNet}

ChannelNet is one of the first deep learning algorithms to be implemented for channel estimation in communication systems. It is a cascaded neural network combining a super-resolution block and a image restoration (IR) module to denoise the channel estimates. The time–frequency response of the channel is considered as two dimensional image. The axes represent time and frequency and each pixel represents the complex channel gain at that time/frequency location. As ReEsNet provides better performance and uses less complexity than Channelnet \cite{b5}, we only consider the ReEsNet network in this paper when comparing simulation results. 

ReEsNet is a residual convolutional neural network shown in Fig.~\ref{ReEsNet} and it can outperform the ChannelNet and approximated LMMSE methods \cite{b10}. It keeps the matrix dimensions of the estimation $\hat{H}_{LS}$ constant and separates the real part and imaginary part of each matrix entry into two channels. It consists of 4 ResBlocks, two convolutional layers and an upsampling module. The first convolutional layer resizes the input features to a large scale of 16 channels. Each of the following 4 ResBlocks is composed of one convolutional layer, one ReLu activation function and one second convolutional layer in series with a skip connection between the input and output. The output from the last ResBlock is processed by the next convolutional layer and added with the output of the first convolutional layer. The upsampling module resizes the input to the output size of $N_f \times N_t \times 2$ by deploying the transposed convolutional layer as shown in Fig.~\ref{ReEsNet}. To our understanding, to avoid `checkerboard artifacts' when resizing the number of channels, the transposed convolutional layer is coupled with a convolutional layer in the upsampling module. Moreover, the residual neural network can mitigate the degradation problem to ensure that deep neural network is fully utilized and a shallow neural network cannot outperform it. 

Unfortunately, the ReEsNet paper does not explain clearly the model settings and instead only refers to \cite{Dumoulin} to explain the transposed convolutional layer operation. In this paper, we consider two variants of ReEsNet which follow very distinct interpretations of this layer and which lead to very different model sizes for the network.
\begin{itemize}
\item \textsl{\textbf{ReEsNet A:}} The kernel size of the transposed convolutional layer is {3 $\times$ 3 $\times$ 16} to match the following convolutional layer \cite{Dumoulin}. Therefore the total number of parameters in the network should be approximately 24,000. 
\item \textsl{\textbf{ReEsNet B:}} The kernel size of transposed convolutional layer is assumed here to be {11 $\times$ 11 $\times$ 16}. Therefore, the total amount of parameters should be approximately 53,000. This value is guessed from the model size quoted for the ReEsNet network in \cite{b7}.
\end{itemize}
Both ReEsNet variants suffer from problems of hard coding when adapting to different pilot patterns. The first two dimensions of the output depends on the size of the input and the hyperparameters of the upsampling module. If the pilot pattern is changed (the size of the input is changed), the hyperparameters of the transposed convolutional layer and the following convolutional layer need to be changed to compensate. This does not provide a flexible design for multiple different 5G pilot patterns and the final performance may degrade.

\subsection{Interpolation-ResNet}

% JST 29/07: updated with short definition of the InterpolationResNet name!

We now introduce a improved neural network structure named Interpolation-ResNet, which deploys bilinear interpolation to resize the output. The network reduces complexity by at least 82\% and can adapt flexibly to different pilot patterns, when compared with the ReEsNet neural network method.

\subsubsection{Interpolation-ResNet network}
%JST 29/07: Figure moved earlier to fit better in paper
\begin{figure}[htbp]
\centering
\subfloat[ReEsNet \label{ReEsNet}]{%
       \includegraphics[width=0.3\linewidth]{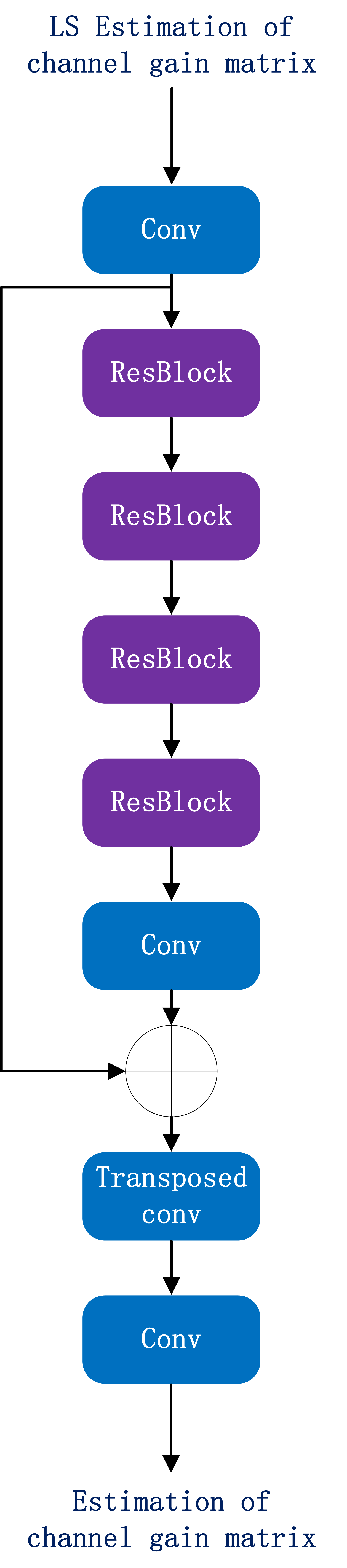}}
\hfill
\subfloat[Interpolation-ResNet\label{Improved neural network}]{%
        \includegraphics[width=0.325\linewidth]{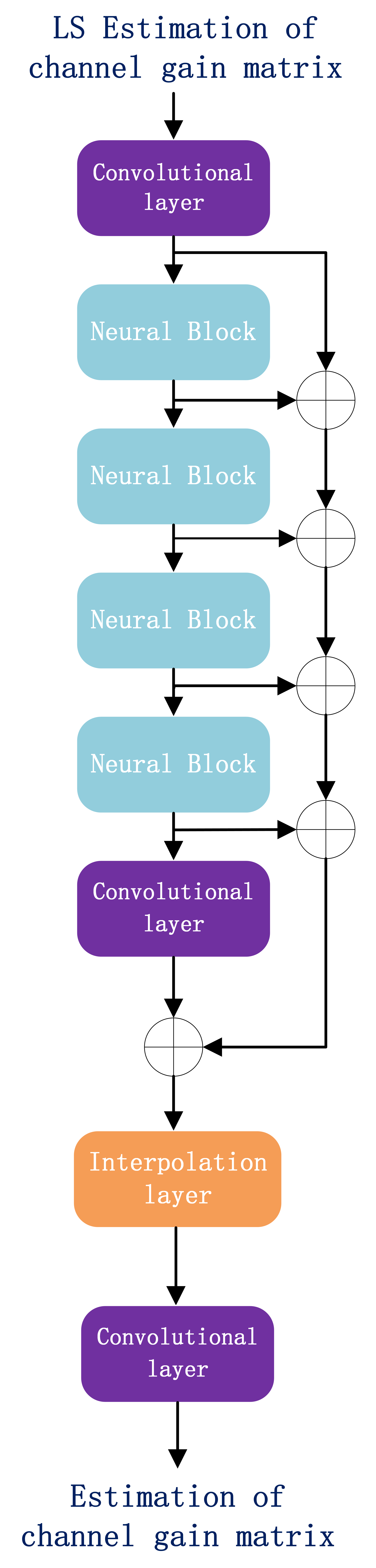}}
\caption{Neural network solutions}
\label{Neural network solutions} 
\end{figure}
The improved neural network architecture is shown in Fig.~\ref{Improved neural network}, which consists of 4 Neural Blocks, 3 convolutional layers and an interpolation layer. The real part and imaginary part of the $\hat{H}_{LS}$ are combined into two channel real matrices as the input of the neural network. The first convolutional layer has \Nf \ filters with the kernel size of {3 $\times$ 3 $\times$ 2} followed by the Res-module, which consists of four neural blocks and one convolutional layer. Each of the following neural blocks consists of one convolutional layer with \Nf \ filters, each corresponding to a kernel size of 3 $\times$ 3 $\times$ \Nf and one ReLu layer and one convolutional layer with \Nf \ filters, and the kernel size of each is 3 $\times$ 3 $\times$ \Nf. In each Neural block, a skip connection adds the input and output of the neural block and forwards it to the next block. The outputs of the first convolutional layer, the convolution layer next to the last neural block and the outputs of each neural block are superimposed by the last addition layer, then the summed result is forwarded to the interpolation layer, which resizes the output using bilinear interpolation. The resized data proceeds to the last layer, which is a convolutional layer with two filters of the kernel size 36 $\times$ 7 $\times$ \Nf. The output of the last convolutional layer is the estimated channel gain matrix (channel 1 for real part and channel 2 for the imaginary part). 
\subsubsection{Bilinear interpolation}

Compared with ReEsNet, Interpolation-ResNet deploys the bilinear interpolation as the upsampling layer. Bilinear interpolation is the two variable version of the linear interpolation, which use the neighbour samples to predict the unknown value at the interpolation positions. The implementation of the bilinear interpolation is illustrated in Fig.~\ref{Bilinear interpolation} and is described as follows: 
\begin{figure}[htbp]
\centerline{\includegraphics[width=0.3\textwidth]{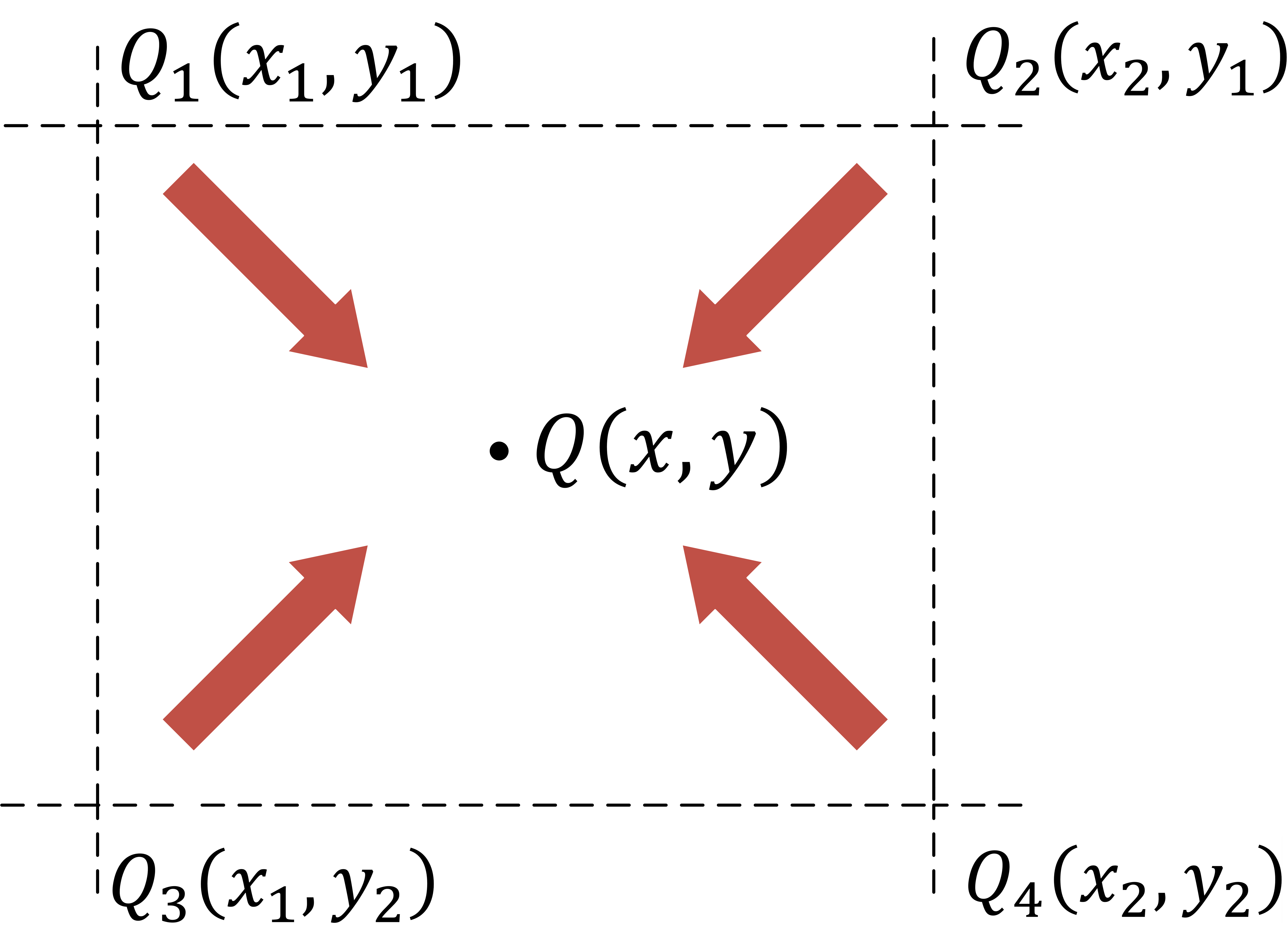}}
\caption{Bilinear interpolation}
\label{Bilinear interpolation}
\end{figure}
\begin{equation}
    f(Q) = \frac{\left[\begin{matrix}x_2 - x & x - x_1\end{matrix}\right]\left[\begin{matrix}f(Q_1) & f(Q_2)\\f(Q_3) & f(Q_4)\end{matrix}\right]\left[\begin{matrix}y_2 - y \\ y - y_1\end{matrix}\right]}{(x_2 - x_1)(y_2 - y_1)}\\
\label{bilinear interpolation}
\end{equation}
Where $f$ denotes the unknown mapping of the samples and $Q1$, $Q2$, $Q3$ and $Q4$ denote the position of the neighbouring samples. Equ.~(\ref{bilinear interpolation}) does not contain any hyperparameters, therefore, the size of bilinear interpolation output completely depends on the output size that we set. Bilinear interpolation resizes the input and passes the result to the last convolutional layer adjacent to the interpolation layer. The overall performance can be improved compared to ReEsNet and the hyperparameters of the neural network proposed do not need to be adjusted the hyperparameters of the transposed convolutional layer for the variable pilot patterns. 

\subsubsection{Complexity reduction and pruning}

For the complexity comparison, Interpolation-ResNet has the amount of parameters of $(36 \times 7 \times 2 \times$ \Nf $)+ (9 \times 3 \times 3 \times (N_\mathrm{filter})^2) +(3 \times 3 \times 2 \times$ \Nf $)+(2+10 \times$ \Nf), compared with the 21,202 parameters used in the ReEsNet A and 52,178 in the ReEsNet B. The complexity of Interpolation-ResNet is shown in Table.~\ref{Complexity of InterpolationResNet} and Interpolation-ResNet-NF means that each convolutional layer of Interpolation-ResNet has \Nf filters.
\begin{table}[htbp]
\caption{Complexity of Interpolation-ResNet}
\begin{center}
\begin{tabular}{|c|c|}
\hline
\textbf{Neural Network}& \textbf{Parameters of Interpolation-ResNet}\\
\hline
\textbf{Interpolation-ResNet-2F}& 1,390\\
\hline
\textbf{Interpolation-ResNet-4F}& 3,426\\
\hline
\textbf{Interpolation-ResNet-6F}& 6,110\\
\hline
\textbf{Interpolation-ResNet-8F}& 9,442\\
\hline
\textbf{Interpolation-ResNet-10F}& 13,422\\
\hline
\end{tabular}
\label{Complexity of InterpolationResNet}
\end{center}
\end{table}
Compared with the ReEsNet B, the computational complexity of Interpolation-ResNet-8F is reduced by 82\%. Neural network pruning \cite{pruning} can also be applied to reduce the complexity of the neural network further. 

Neural network pruning is a method that removes insignificant weights or neurons, which allows a trade-off the performance and efficiency of the model. The absolute value of the weights are sorted to evaluate and the importance of each one is ranked. The smallest weights are set to 0 according to the desired pruning rate to reduce the amount of the parameters in the model. For example, a 10\% pruning rate means that the smallest 10\% of the neural connections are removed. It should be noted that the neural network is simplified with a concomitant performance loss. If the performance loss is acceptable, neural network pruning will be implemented to obtain a more sparse version of the trained neural network. Pruning results are shown in Section \ref{Prune}. 

\section{Simulation results}
\label{Simulation results}

In this paper, we consider a downlink scenario in a single-input and single-output (SISO) system where the user has a randomly selected mobile speed from 0km/h to 50km/h. The propagation channel models deployed are the 3GPP Extended Pedestrian A model (EPA), Extended Vehicular A model (EVA) and Extended Typical Urban model (ETU) \cite{b11}. 

\subsection{System settings}

The detailed system settings are introduced in the Table \ref{Baseband Parameter}. Two pilot patterns are considered for the results in this paper:\\
\textsl{\textbf{Default:}} The 1\textsuperscript{st} and 13\textsuperscript{th} OFDM symbols of each frame are selected to be pilot symbols \cite{b12}. For the first pilot symbols, the pilot subcarriers start from the 1\textsuperscript{st} subcarrier and are spaced by 3 subcarriers. For the second pilot symbols, the pilot subcarriers start from the 2\textsuperscript{st} subcarrier and are also spaced by 3.\\
\textsl{\textbf{Alternate:}} The 1\textsuperscript{st}, 5\textsuperscript{th}, 9\textsuperscript{th} and 13\textsuperscript{th} OFDM symbols of each frame are selected to be pilot symbols \cite{b12}. For each pilot symbol, the pilot subcarriers start from the 1\textsuperscript{st}, 2\textsuperscript{nd}, 4\textsuperscript{th} and 6\textsuperscript{th} subcarrier and are spaced by 6 subcarriers.\\
Therefore, 48 pilots are used for one frame in both options. The \textsl{default} pilot option is used unless otherwise stated.
% JST 29/07 - Table moved later to help fit in the text
\begin{table}[htbp]
\caption{Baseband Parameters}
\begin{center}
\begin{tabular}{|c|c|}
\hline
\textbf{Parameter}& \textbf{Value}\\
\hline
\textbf{Pilot Subcarriers - default (alternate)}& 24 (12)\\
\hline
\textbf{Pilot Symbols - default (alternate) }& 2 (4) \\
\hline
\textbf{Number of deployed subcarriers}& 72\\
\hline
\textbf{CP Length}& 16\\
\hline
\textbf{Bandwidth}& 1.08MHz\\
\hline
\textbf{Carrier frequency}& 2.1GHz\\
\hline
\textbf{Subcarrier Spacing}& 15kHz\\
\hline
\textbf{Number of frame per slot}& 1\\
\hline
\textbf{Number of OFDM symbols per slot}& 14\\
\hline
\end{tabular}
\label{Baseband Parameter}
\end{center}
\end{table}
\subsection{Parameters for ReEsNet/Interpolation-ResNet Training}

% JST 29/07 - fixed problem with non printing tilde (\sim in math mode)
The training dataset is generated with the SNR range from 0dB to 20dB and maximum Doppler frequency shift selected from 0Hz to 97Hz (corresponding to 0km/h $\sim$ 50km/h, medium speed) in the EPA downlink scenario. The training dataset is comprised of 20,000 channel realization samples for each integer SNR to achieve sufficient performance. Therefore, the total size of the training dataset is 100,000 samples, 95\% for training and 5\% for validation. Interpolation-ResNet and ReEsNet use the exactly same training dataset. The detailed training parameters are introduced in Table \ref{Training Parameter}. 

\subsection{MSE performance over SNR}

\begin{figure}[htbp]
\centerline{\includegraphics[width=0.9\linewidth]{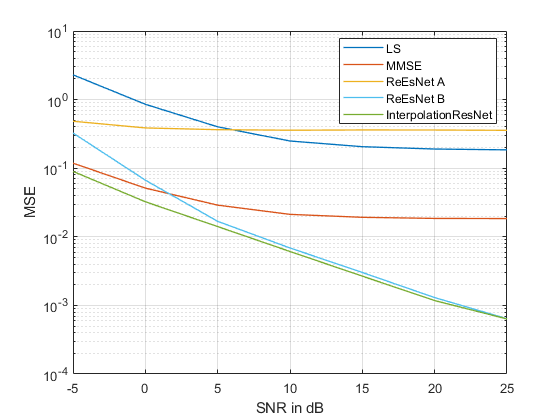}}
% JST 29/07 - updated to more meaningful caption
\caption{Performance of the estimators for the EPA channel}
\label{MSE performance (EPA)}
\end{figure}

MSE is the performance metric used in this paper to evaluate the error between the desired and estimated channels. It is defined as 
% JST 29/07 - updated this equation to be more consistent with the paper notation.
\begin{equation*}
    MSE(\hat{H}, H) = \frac{1}{N_f N_t}\sum_{i=1}^{N_f} \sum_{j=1}^{N_t} E\left\{\left|\hat{H}_{ij} - H_{ij}\right|^2\right\}
\end{equation*}
Where $H_{ij}$ is the exact channel at subcarrier $i$ and OFDM symbol $j$ and $\hat{H}_{ij}$ is its estimated value. To show that the proposed method is valid in an extended range, the test dataset is generated with extended SNR range from -5dB to 25dB and maximum Doppler frequency shift from 0Hz to 97Hz in the EPA downlink scenario. Each SNR is tested with 5000 independent channel realizations to average over the Monte Carlo effects. 
\begin{table}[htbp]
\caption{Training Parameters}
\begin{center}
\begin{tabular}{|c|c|}
\hline
\textbf{Parameter}& \textbf{Value}\\
\hline
\textbf{Optimizer}& Adam\\
\hline
\textbf{Loss function}& Mean squared error (MSE)\\
\hline
\textbf{Maximum epoch}& 100\\
\hline
\textbf{Initial learning rate}& 0.001\\
\hline
\textbf{Drop period for learning rate}& 20\\
\hline
\textbf{Drop factor for learning rate}& 0.5\\
\hline
\textbf{Minibatch size}& 128\\
\hline
\textbf{L2 regularization}& 0.001\\
\hline
\end{tabular}
\label{Training Parameter}
\end{center}
\end{table}
Fig.~\ref{MSE performance (EPA)} compares the MSE results of conventional methods and neural network solutions when the EPA channel model is implemented for testing. It can be seen that Interpolation-ResNet-8F outperforms the LS method and ReEsNet A, while achieving similar performance to ReEsNet B. Interpolation-ResNet-8F can also outperform the MMSE method used in this paper. The reason is that, the exact channel gain matrix of the whole frame is used as the training label for Interpolation-ResNet-8F and ReEsNet(A, B). By contrast, the MMSE method only has access to the channel gain matrix at the pilot position and the neural network is trained on the EPA channel model. 

\subsection{Generalization capability}

To prove that the Interpolation-ResNet neural network is capable of generalizing to the extended propagation channel models, Interpolation-ResNet-8F and ReEsNet B (trained with EPA as mentioned) are both tested on the EPA, EVA and ETU channels over the extended SNR range from -5dB to 25dB and with Doppler frequency shifts from 0Hz to 97Hz (correspondingly from 0km/h to 50km/h, medium speed). Each result of the deployed channel models is simulated with 35,000 (5000 $\times$ 7) independent channel realizations to average over the Monte Carlo effects. 
\begin{figure}[htbp]
\centering
\subfloat[Testing different channel models \label{Performance of generalization}]{%
        \includegraphics[width=0.5\linewidth]{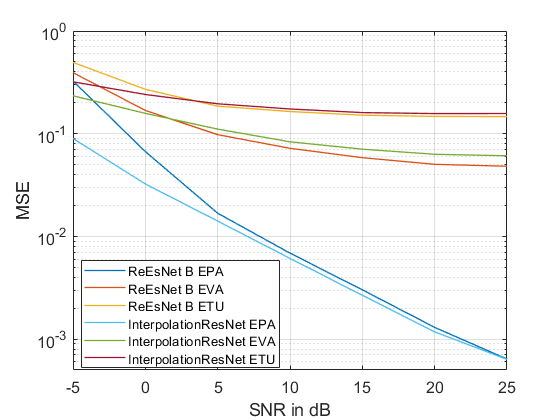}}
\hfill
\subfloat[Testing \textsl{alternate} pilot option \label{Test with another pilot pattern on the EVA channel}]{%
       \includegraphics[width=0.5\linewidth]{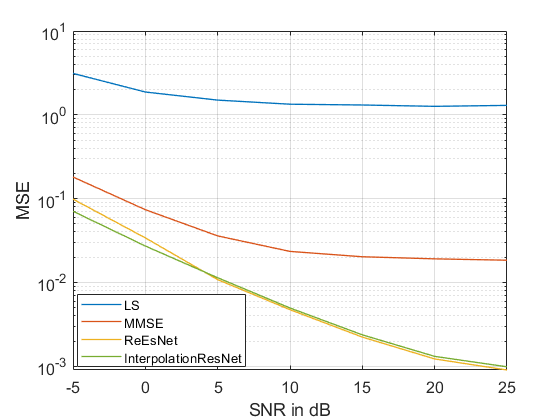}}
\caption{Generalization capability of the networks to different channel models and pilot patterns}
\label{Generalization capability} 
\end{figure}
From Fig.~\ref{Performance of generalization}, the MSE curves flatten out to constant values with increasing SNR when the EVA and ETU channels are tested. Interpolation-ResNet-8F only outperforms ReEsNet B when tested on the channel it was trained for, namely EPA. When tested on the EVA and ETU channel, Interpolation-ResNet-8F only outperforms ReEsNet B in the low SNR range of -5dB to 0dB, which suggests that the generalization to different channel models of ReEsNet B is slightly better than Interpolation-ResNet-8F. However, both methods' generalization capability is clearly insufficient in this example. The training dataset mentioned is generated based on the EPA channel, therefore, the neural network solutions perform best when tested on the same EPA channel. The power delay profile of the ETU channel is significantly different from the EPA channel, therefore, the MSE results for the ETU channel are unacceptably high for both networks. It would be better to use a wider range of channels for training both networks.

We also test the Interpolation-ResNet-8F and ReEsNet B with the \textsl{alternate} pilot pattern, an sparse pilot pattern using 4 pilot symbols which also contains 48 pilots in total, to show the flexibility and adaptability to different pilot patterns.
% JST 29/07 - this text now moved to start of this section.
Interpolation-ResNet-8F and ReEsNet B are retrained by the dataset generated on the EPA channel over the SNR range from 0dB to 20dB and maximum Doppler frequency shift from 0Hz to 97Hz (50 epochs). 

% JST 29/07 - updated this par
Fig.~\ref{Test with another pilot pattern on the EVA channel} compares the MSE results of conventional methods and neural network solutions when testing with the \textsl{alternate} pilot pattern on the EPA channel over the extended SNR range from -5dB to 25dB with maximum Doppler frequency shifts from 0Hz to 97Hz. Compare with ReEsNet, the MSE performance of Interpolation-ResNet-8F is almost same. However, the amount of the parameters are reduced by 82\% and the hyperparameters of the ReEsNet neural network need to be adjusted to keep the output size constant while Interpolation-ResNet does not. 

\subsection{Complexity reduction}
\label{Prune}

\begin{figure}[htbp]
\centering
\subfloat[Interpolation-ResNets performance \label{InterpolationResNets}]{%
       \includegraphics[width=0.5\linewidth]{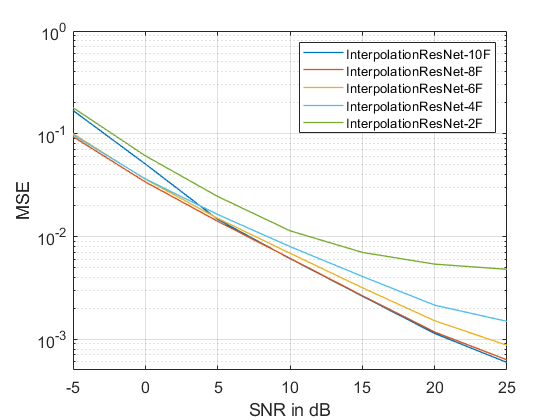}}
\hfill
\subfloat[InterpolationResNet-8F pruning \label{InterpolationResNet pruning}]{%
        \includegraphics[width=0.5\linewidth]{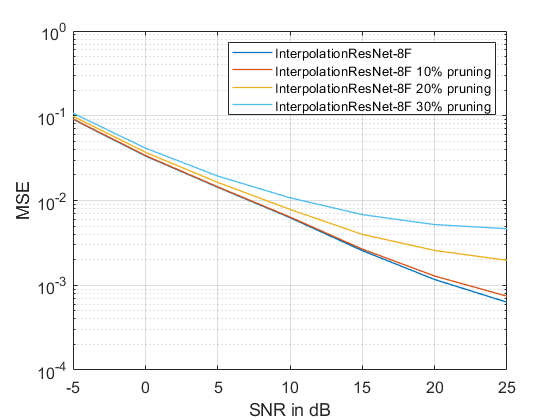}}
\caption{Complexity reduction}
\label{Neural network pruning} 
\end{figure}

To reduce the amount of the parameters, the performance of Interpolation-ResNet-(10F, 8F, 6F, 4F, 2F) are shown in Figure.~\ref{InterpolationResNets}. The test dataset is generated with the extended SNR range from -5dB to 25dB and maximum Doppler frequency shift from 0Hz to 97Hz in the EPA downlink scenario. From the results, the overall MSE plots of Interpolation-ResNet-(10F, 8F, 6F, 4F) are almost same. Interpolation-ResNet-2F is impacted severely but it only has 1,390 parameters, which is a reduction of 97\% in total parameters when compared with ReEsNet B. 

To reduce the redundant computations in the neural network, neural network pruning is applied for Interpolation-ResNet-8F for the \textsl{default} pilot pattern and test with the same dataset. Fig.~\ref{InterpolationResNet pruning} shows the pruning results of Interpolation-ResNet-8F by the pruning rates of 0\%, 10\%, 20\% and 30\%. From the results, Fig.~\ref{InterpolationResNet pruning} demonstrates that a pruning rate above 10\% have a more severe impact on the overall performance of Interpolation-ResNet-8F at the high SNR range. It proves that Interpolation-ResNet is a optimised and low complexity network. 

\section{Conclusion}
\label{Conclusion}

We propose a general residual neural network method called Interpolation-ResNet with low complexity for downlink channel estimation. We investigate the generalization capability of the proposed neural network on the extended SNR range, different propagation channel models and different downlink pilot patterns. From the simulation results, the performance of Interpolation-ResNet is better than the baseline methods when tested on the EPA channel. It also reduces complexity by 82\% compared to the ReEsNet-B method. Moreover, the limited generalization capability and adaptability of Interpolation-ResNet is also shown for different channel models and different downlink pilot patterns. Neural network pruning is also evaluated as the method to further reduce the neural network complexity. 

\section{Acknowledgement}

The authors gratefully acknowledge the funding of this research by Huawei.

\vspace{12pt}

\end{document}